\documentclass[aps,prb,twocolumn,groupedaddress,showpacs,letterpaper]{revtex4}

\usepackage[dvips]{graphics}

\begin{document}
\title{Dynamics of topological defects in a 2D magnetic domain stripe pattern}
\author{N. Abu-Libdeh} \author{D. Venus}
\email{venus@physics.mcmaster.ca} 
\affiliation{Dept. of Physics and
Astronomy, McMaster University, Hamilton Ontario, Canada}
\date{\today}

\begin{abstract}
Two dimensional magnetic films with perpendicular magnetization spontaneously form magnetic domain patterns that evolve or undergo symmetry transformations as a function of temperature.  When the system is driven from equilibrium by a rapid change in temperature, topological pattern defects are the elementary pattern excitations that affect this evolution.  An elastic continuum model is adapted to describe how a metastable population of topological defects alters the domain density and the magnetic susceptibility of the ``stripe'' magnetic domain pattern.  Temporal changes in the susceptibility are interpreted using a dynamical equation describing the defect population. Recent experiments provide a quantitative verification of the model, and illustrate the use of the magnetic susceptibility to follow the population dynamics of topological defects in this system, and its potential role in investigating a pattern melting phase transition.
\end{abstract}
\maketitle

\section{Introduction}

The strong thermal fluctuations that occur in two dimensional magnetic systems have a dramatic effect on long range ordering.\cite{chaikin}  Even when the presence of anisotropies allows an ordered state,\cite{bander} it is rather fragile and can be complicated by the presence of competing interactions.  For example, many two dimensional systems with short range attractive interactions and long range repulsive interactions spontaneously form mesoscopic patterns that destabilize the uniformly ordered state.\cite{sagui}  Examples are the structure of Langmuir-Blodgett films\cite{sikes}, of adsorbates on surfaces,\cite{vanderbilt} and the spin/charge structure of some high temperature superconductors.\cite{kivelson}  Topological defects in the pattern play an important role in their dynamics.\cite{chaikin,sagui}

Ultrathin magnetic films magnetized normal to the plane of the film are an example of a two dimensional system with competing interactions that spontaneously forms patterns.\cite{skomski}  At low temperature, a stripe pattern of up and down magnetization domains develops, with the width of the stripes exponentially dependent upon the temperature.  Experimental studies using various magnetic microscopies have observed topological defects in the pattern,\cite{allenspach} either in the form of domain stripe segments that end with a rounded endcap and an accompanying distortion of the spacing and curvature of adjascent domains, or as bound pairs of these dislocations.  Numerical simulations show similar features.\cite{stoycheva}  Attention thus far has been focussed on the role of these defects in the ``melting'' of the domain pattern\cite{vaterlaus,won} at a Kosterlitz-Thouless transition\cite{chaikin,kosterlitz,kashuba,abanov} through the proliferation of topological defects.  The quantitative study and confirmation of the melting transition, and its relationship to either a spin-rorientation transition, or to the creation of the paramagnetic state, remains an experimental challenge that is complicated by the effects of domain wall pinning.

A separate, but related, issue is the role of topological defects in moving the domain pattern towards equilibrium.  Changes in temperature, for example, imply an exponential change in the stripe domain density.  Since domain stripes cannot appear fully-formed, domain growth and removal occurs by the motion of dislocations.\cite{saratz,portmann}  The creation of a new domain stripe is complete when two dislocations of the same domain type meet and join. On the other hand, dislocations of the opposite domain type originating at separated points will not meet ``head on'', but will be separated laterally by one stripe width, forming  a bound dislocation pair.  This arrangement is metastable because it does not disrupt the equilibrium stripe density or create a net moment.  The larger scale domain rearrangements required to annihilate the pair of dislocations require statistically rare events where many local fluctuations are correlated.  In numerical simulations of 2D patterned systems, a distribution of topological defects typically remains long after the mean domain density has relaxed to its equilibrium value.\cite{bromley,cannas}  Many patterned systems then relax on a very long time scale through the pairwise annihilation of the dislocations.\cite{sagui}

The relaxation of topological defects has proven difficult to study using local imaging.  Imaging techniques can record the presence of a small number of pattern defects, each of which is very unlikely to be annihilated.  If an annihilation event occurs, it happens too quickly to be followed.\cite{portmann2}  Similarly, numerical simulations with a time resolution small enough to reproduce the mechanics of an annihilation event must run for a very long time.  This mismatch in time scales can be overcome by using experimental techniques that are not local, but average over a large sample and follow the relaxation dynamics of a population of topological defects.  An example is measurements of the magnetic susceptibility.

This article adapts analytic theories for the equilibrium of a magnetic stripe domain system,\cite{kashuba,abanov} to include the domain energy contained in a metastable distribution of topological defects. By including the defect energy, the domain density and magnetic susceptibility are altered from the equilibrium result, and predict changes in the shape and temporal relaxation of the experimentally measured susceptibility.  These predictions are compared to recent experiments\cite{libdeh1,libdeh2} measuring the magnetic susceptibility of perpendicularly magnetized xML Fe/2ML Ni/W(110) films that are heated at a constant rate.  The excellent qualtitative and quantitative agreement between the experiments and the model demonstrates that the population dynamics of the defects can be followed experimentally, and shows that relaxation times and activation energies for the annihilation of topological defects can be extracted from the data.

\section{Magnetic response due to topological defects}
\subsection{The magnetic susceptibility due to domain wall motion}

An analytic treatment of the domain patterns of a two dimensional perpendicularly magnetized film, in the continuum approximation, has been given by Kashuba \emph{et al.}\cite{kashuba} and by Abanov \emph{et al.}\cite{abanov}.  The results are very slightly modified here to account for a film with a thickness of $d=Nb$, where $N$ is the number of monolayers with lattice constant $b$.  A film of area $L_x \times L_y$ with a regular pattern of straight-edged stripes of width $L=1/n$ has an areal energy density $E_0$ due to domains.
\begin{equation}
\label{E_0}
E_0= E_WNn -4\Omega N^2 n \ln[{\frac{2}{\pi \ell n}}],
\end{equation}
$\Omega=\frac{\mu_0}{2}\frac{(g \mu_B S)^2}{a^4}$ is a constant that sets the scale of the dipole energy ($a$ is the in-plane lattice constant). $E_W$ is the energy per unit length of a straight domain wall, and $\ell$ is the width of the domain wall. The exchange stiffness is $\mathcal{A}=(zJS^2)/2$, with $z$ the number of nearest neighbours, and the effective areal anisotropy is 
\begin{equation}
\label{Keff}
K_{eff}(T)=\frac{K_S(T)}{N}-\frac{\Omega}{b},
\end{equation}
which includes both the surface anisotropy $K_S(T)$ and the short range dipole, or demagnetization, energy.

In terms of these areal definitions $E_W=4\sqrt{\mathcal{A}K_{eff}}$ and $\ell = \pi \sqrt{\frac{\mathcal{A}}{K_{eff}}}$.\cite{chikazumi}  The second term in eq.(\ref{E_0}) represents the reduction in dipole energy due to forming uniform stripe domains, and the first term gives the accompanying increase in domain wall energy.  Balancing these factors gives an equilibrium domain density $n_{eq}(T)$, where
\begin{equation}
\label{neq}
n_{eq}(T) = \frac{2}{\pi \ell} \exp(-\frac{E_W(T)}{4\Omega N}-1).
\end{equation}
The variation of $K_{eff}(T)$ in $E_W$ drives the exponential change in domain density with temperature.

Applying a small, perpendicular field causes a net moment as the domain widths are perturbed by lateral movement of the domain walls.  In equilibrium, this leads to a dc magnetic susceptibility
\begin{equation}
\label{chi0}
\chi_{eq}(T)=\frac{2}{\pi^2 d n_{eq}(T)} \approx A_0\exp[{-\kappa_0 T}].
\end{equation}
The phenomenological constants $A_0$ and $\kappa_0$ have proven to give an excellent representation of experimental data.\cite{venus1,venus2}   A linear expansion about temperature $T_0$ gives
\begin{equation}
\label{lnA}
\ln A_0 = \frac{1}{4\Omega N} E_W(T_0),
\end{equation}
\begin{equation}
\label{kappa}
\kappa_0=\frac{1}{4\Omega N} \frac{\partial E_W}{\partial T}|_{T=T_0}.
\end{equation}
If an ac field of angular frequency $\omega$ is used, then, in the relaxation approximation, the susceptibility is modified by a dynamical factor:\cite{venus1}
\begin{equation}
\label{ac}
\chi(T) = \frac{1-i\omega \tau_p(T)}{1+\omega^2 \tau_p^2(T)}\ \chi_{eq}(T).
\end{equation}
$\tau_p(T)$ is the relaxation time for the pinning of domain wall segments at structural inhomogeneities.  It has activation energy $E_p$ and fundamental time scale $\tau_{0,p}$, such that $\tau_p(T) = \tau_{0,p}\exp[{\frac{E_p}{kT}}]$. 

\subsection{Contribution of topological pattern defects to the susceptibility}
The existence of metastable topological defects in the domain pattern will alter the domain energy function in eq.(\ref{E_0}), and through this the metastable domain density.  This can be illustrated by considering the additional domain energy due to a simple idealized dislocation, represented by a semicircular endcap of radius $R=L/2$ that terminates a stripe domain segment. Using cylindrical co-ordinates relative to an origin at the centre of a circular domain wall with $R>>\ell$, the total energy of the curved domain wall endcap can be written as\cite{darby}
\begin{equation}
\label{Ewall}
E_{wall}=N\pi R E_W \sqrt{1+\frac{\ell^2}{\pi^2R^2}},
\end{equation}
where $E_W$ and $\ell$ are the quantities defined for a straight domain wall in the previous section.

The important point is that both the energy due to the additional length of wall in the dislocation, and due to its curvature (the second term in the square root in eq.(\ref{Ewall})), are proportional to $E_W$.  Thus, a calculation of the dislocation energy using a realistic geometry where nearby domain walls are distorted would also have terms due to additional domain wall length and curvature that are proportional to $E_W$.  An additional contribution to the energy that is ultimately related to the resulting perturbation of the dipole energy, is the change to the effective compressional energy when the domains are not uniformly spaced near the dislocation.  However, as Abanov \emph{et al.}\cite{abanov} note, in local equilibrium this energy is also determined by the additional domain wall length.  Therefore, the energy density of the domain pattern, eq.(\ref{E_0}) contains an additional term proportional to $E_W$ when dislocations are present.

Kashuba \emph{et al.}\cite{kashuba} give a general expression for the additional domain pattern elastic energy when the domain walls are not uniform and straight, but are described by the path $u(x,y)$.  Although they concentrate on the equilibrium state at finite temperature, and the Kosterlitz-Thouless transition, the elastic energy of metastable dislocations can also be evaluated. The additional elastic energy per unit area due to domain wall curvature is
\begin{equation}
\label{curve}
E_{curve}= \frac{N}{L_xL_y}\Omega L \int d^2r \ \frac{1}{2}(\frac{\partial ^2 u}{\partial y^2})^2 \equiv N\Omega L \frac{\beta(L)}{L_xL_y}.
\end{equation}
The additional energy density due to changes in domain wall length and spacing (compressibility) are given by
\begin{equation}
\label{comp}
E_{comp} = \frac{N}{L_xL_y}\frac{\Omega}{L}\int d^2r\ [\frac{\partial u}{\partial x}+\frac{1}{2}(\frac{\partial u}{\partial y})^2]^2 \equiv \frac{N\Omega}{L}\frac{\alpha(L)}{L_xL_y}.
\end{equation}

For a localized topological defect, the region where the integrals $\alpha (L)$ and $\beta (L)$ are non-zero will be restricted.  This restricted range is denoted by a superscript ``0'' on the integrals . If there are $Q$ such non-overlapping regions containing topological defects, then the energy density in eq.(\ref{E_0}) becomes
\begin{equation}
\label{Qdyn}
E = E_0 + \frac{N\Omega}{L}\frac{Q}{L_xL_y}\alpha^0(L) +N\Omega L \frac{Q}{L_xL_y}\beta^0(L).
\end{equation}

There are now two possibilities.  The elastic energies may arise from domain geometries that do not scale quasi-statically with $L$.  An example might be a region of dimension $L_0$ where the domain stripes are oriented along a different axis.  Another is a defect that is anchored to a structural defect of size $L_0$. This circumstance is treated in the appendix. 

Alternatively, the defect geometry may scale with $L$. For example, in the case of a simple dislocation or a bound pair of dislocations, the domain density changes on a much faster time scale than the topological defects annihilate.  Since the cores of these defects are the terminations of stripe segments, their size will change quasi-statically with the stripe width $L$.  The integrals become dimensionless when all distances are scaled by $L$, giving $\alpha^0(L)\rightarrow L^2 \alpha^0$ and $\beta^0(L)\rightarrow \beta^0$.  
Furthermore, when the defects scale with $L$, it is useful to express the density of dislocation pairs, $Q/(L_xL_y)$ relative to the scale of the domain pattern as well.  That is, if $N_xN_y=(L_x/L)(L_y/L)$ is the number of cells of size $L^2$ in the sample, then the fraction $q$ of these cells containing a defect is $q=Q/(N_xN_y)$.  Collecting all these definitions together gives
\begin{equation}
E = E_0 + N\Omega(\alpha^0 +\beta^0)\frac{q}{L}.
\end{equation}

Finally, the dimensionless integral $(\alpha^0 +\beta^0)$ must be proportional to $E_W$, in agreement with general considerations exemplified by the very simplified model treated at the beginning of this section.  This means that it must scale as $\gamma E_W/\Omega$.  The dimensionless factor $\gamma$ depends upon the detailed positions of the domain walls in the topological defect, and must be calculated within a specific model.  Incorporating these results into eq.(\ref{E_0}) for the energy density of the domain pattern yields the energy density in the presence of a fraction $q$ of the region occupied by topological defects:
\begin{equation}
\label{E_final}
E= (1+\gamma q)E_WNn -4\Omega N^2 n \ln{[\frac{2}{\pi \ell n}]}.
\end{equation} 
In the presence of the topological defects the metastable domain density is
\begin{equation}
\label{nmeta}
n_{ms}= \frac{2}{\pi \ell}\exp[-(1+\gamma q)\frac{E_W}{4\Omega N}-1],
\end{equation}
and there is an accompanying change in the susceptibility so that $\chi \sim 1/n_{ms}$.  According to these results, the phenomenological quantities $\kappa$ and $A$ that describe the susceptibility in the presence of topological defects, are related to those in eq.(\ref{kappa}) and (\ref{lnA}) by
\begin{equation}
\label{dk}
\kappa=(1+\gamma q)\kappa_0,
\end{equation}
\begin{equation}
\label{dA}
\ln A=(1+\gamma q)\ln A_0.
\end{equation}
The presence of topological defects therefore increases the  domain density, and increases $\kappa$. This effectively increases the magnetic stiffness of the system.  The increase in $\kappa$ decreases the temperature, $T_{pk}$, at which the susceptibility is maximum. An estimate of this effect can be calculated using eq.(\ref{ac}). $T_{pk}$ can be defined implicitly by the condition $\frac{\partial \chi(T)}{\partial T}|_{T=T_{pk}}=0$.  Differentials then yield
\begin{equation}
\label{shift}
\frac{\Delta \kappa}{\kappa_0}= -2[1+\frac{E_p}{kT_{pk}}-\frac{\kappa_0 T_{pk}}{2}] \frac{\Delta T_{pk}}{T_{pk,0}}.
\end{equation}
Finally, it is clear that the changes in $\kappa$ and $\ln A$ in eq.(\ref{dk}) and (\ref{dA}) are related.  For susceptibilities measured with different peak temperatures that arise from different defect densities $q$, a plot of $\ln A$ vs. $\kappa$ will be linear.  According to eq.(\ref{lnA}) and (\ref{kappa}), the slope of the plot will be
\begin{equation}
\label{slope}
\frac{\ln A_0}{\kappa_0}=2\frac{K_{eff}(T_0)}{\frac{\partial K_{eff}(T)}{\partial T}|_{T=T_0}}.
\end{equation}

\subsection{Dynamics of the density of topological defects}

These consideration illustrate that a population of topological defects in an ultrathin film alters the magnetic susceptibility. The population dynamics of the defects should therefore be reflected in temporal changes in the susceptibility.  For clarity of discussion, consider the density of bound dislocation pairs that is produced by the exponential increase in domain density with temperature.  When the magnetic susceptibility is measured at a constant rate of heating, $R$, the number of bound dislocation pairs will evolve through two processes.  First, the bound pairs will annihilate with some average relaxation time $\tau_{an}$ that can be represented by an Arrhenius law.  Second, new bound pairs will be created as the stripe density increases upon heating.  This is because the stripe density increases by the growth of existing dislocations towards each other, by the nucleation of segments that grow by extending the dislocations located at either end, or by ``budding'' of new domain branches from the edge of existing stripes.\cite{portmann}   Since the growth of a stripe domain begins at well separated points, two advancing dislocations of the same type do not always meet head-to-head to form a continuous stripe.  In some fraction of the cases, dislocations of opposite type will meet with a lateral displacement and form instead a bound pair. This statistical creation process will scale with the number of stripe domains that are grown, or equivalently, with the number of $L\times L$ cells in the film.  The evolution can then be written as
\begin{equation}
\label{diffeq0}
\frac{\partial Q}{\partial t}=-\frac{Q}{\tau_{an}}+\epsilon \frac{\partial}{\partial t}N_xN_y ,
\end{equation}
where $\epsilon$ is the fraction of the $L\times L$ areas where a dislocation pair is generated by domain growth.  Converting from the number of dislocation pairs, $Q$, to their fractional density, $q$, gives
\begin{equation}
\frac{\partial q}{\partial t}=-2(q-\epsilon)\frac{1}{n_{ms}}\frac{\partial n_{ms}}{\partial t} -\frac{q}{\tau_{an}}.
\end{equation}

The differential with respect to time can be converted to one with respect to temperature using the constant heating rate $R$. This is allowed because $\tau_{an}\not=\tau_{an}(L)$.
Then, eq.(\ref{nmeta}) and (\ref{dk}) give
\begin{equation}
\frac{1}{n_{ms}}\frac{\partial n_{ms}}{\partial T}=\frac{\partial}{\partial T}\ln n_{ms} = \kappa = (1+\gamma q)\kappa_0.
\end{equation}
Substituting this gives the differential equation
\begin{equation}
\label{diffeq}
\frac{1}{2\kappa_0}\frac{\partial q}{\partial T}=-[(1-\epsilon \gamma)+\frac{1}{2\kappa_0 R\tau_{an}}]q -\gamma q^2 + \epsilon.
\end{equation}
Finally, eq.(\ref{dk}) can be used to convert to an equation for the relative change $\Delta\kappa = \kappa -\kappa_0$:
\begin{equation}
\label{diffeqk}
\frac{1}{2\kappa_0}\frac{\partial}{\partial T}(\frac{\Delta\kappa}{\kappa_0})=-[(1-\epsilon \gamma)+\frac{1}{2\kappa_0 R\tau_{an}}](\frac{\Delta\kappa}{\kappa_0}) - (\frac{\Delta\kappa}{\kappa_0})^2 + \gamma\epsilon.
\end{equation}
Since $\kappa(T,R)$ depends only on the product $\gamma\epsilon$, the number of independent variables is reduced.  This equation must be integrated numerically.  However, the limiting behaviours are simple.  If the heating rate is sufficiently high, then annihilation is ineffective and $q\rightarrow \epsilon$ while $\kappa \rightarrow (1+\gamma \epsilon) \kappa_0$.  If the heating rate is very slow, so that annihilation dominates the dynamics, then $q\rightarrow 0$ and $\kappa \rightarrow \kappa_0$.

\section{Comparison to experiments}

Recent experiments\cite{libdeh1,libdeh2} measuring the susceptibility of Fe/2 ML Ni/W(110) films provide an opportunity to test for the presence of a density of topological defects through magnetization dynamics. The measurements involved cooling the films at a constant rate of 0.1 K/s, and then measuring the low-frequency ac susceptibility as a function of temperature, while heating at a constant rate R.  A series of measurements were made on the same film for a range of heating rates, and films of three different Fe thicknesses were investigated.  Experimental details can be found in the original papers.

The main experimental finding is that the measured susceptibility depends on the heating rate through a relaxation process.  The whole of the susceptibility curve relaxes to higher temperature as the heating rate is reduced.  The relaxation is characterized by an Arrhenius law when the peak temperature, $T_{pk}(R)$, of the susceptibility is plotted as a function of the number of thermally activated time constants, $\tau_r$, that elapse during the time the film is heated to the temperature $T_{pk}(R)$.\cite{libdeh1}  The relaxation is very slow, with a fundamental time scale of $\tau_{0,r}$=0.7 s, and an activation energy of $E_r=1560 K$.  This relaxation time is at least three orders of magnitude slower than the relaxation of the domain density to its equilibrium value.  According to ref.(\onlinecite{libdeh2}), departures from the equilibrium domain density on the experimental time scale are detected only for the fastest heating rates of the thicker films.  The current analysis focuses 1.5 ML Fe films, where these effects are absent.
\begin{figure} 
\scalebox{.4}{\includegraphics{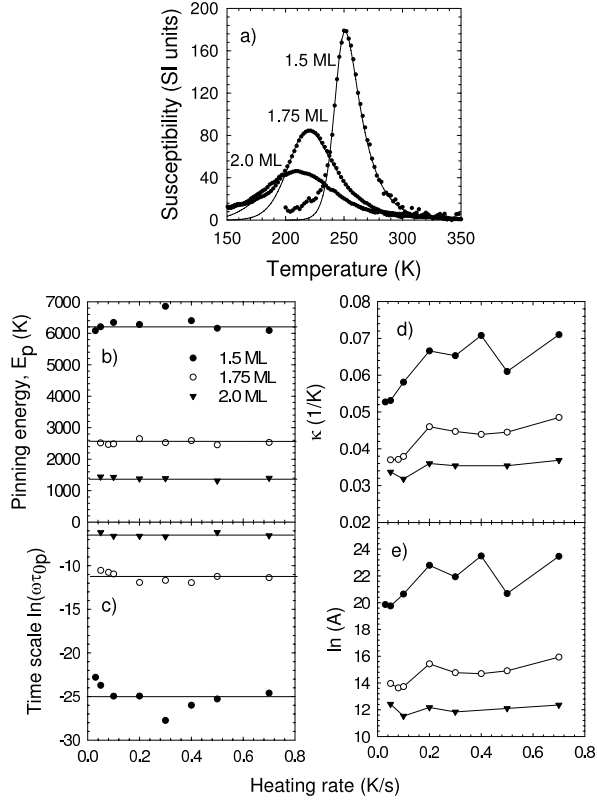}} 
\caption{Magnetic susceptibility of xML Fe/2ML Ni/W(110) first reported in ref.(\onlinecite{libdeh2}).  a) Representative data for a heating rate of 0.3 K/s for three Fe film thicknesses.  The solid line is the result of fitting eq.(\ref{ac}) to the data to obtain parameters presented in the other sections of the figure.  b) The pinning energy $E_p$ as a function of heating rate. c) The characteristic relaxation time as a function of heating rate. d) The parameter $\kappa$ as defined in eq.(\ref{kappa}), and e) the parameter $\ln A$ as defined in eq.(\ref{lnA}).}
\end{figure}

The original susceptibility traces can be viewed in ref.(\onlinecite{libdeh2}).  Here, instead, fig. 1 presents the parameters $E_p,\ \ln(\omega\tau_{0,p}),\ \ln A$ and $\kappa$ derived from fitting eq.(\ref{ac}) to the susceptibility. Fig. 1a shows representative fits to the data for three film thicknesses for the heating rate $R$=0.3 K/s.  Fig. 1b and 1c give the parameters that characterize the domain wall pinning by structural inhomogeneities, as a function of $R$.  These are essentially unaffected by changing the heating rate, as would be expected.  Fig. 1d and 1e presents the parameters $\kappa$ and $\ln A$ that characterize the susceptibility and domain density at higher temperature where pinning is not effective.  These  depend upon the heating rate in a correlated fashion, with $\kappa$ increasing as the heating rate increases, consistent with the presence of topological defects.
\begin{figure} 
\scalebox{.4}{\includegraphics{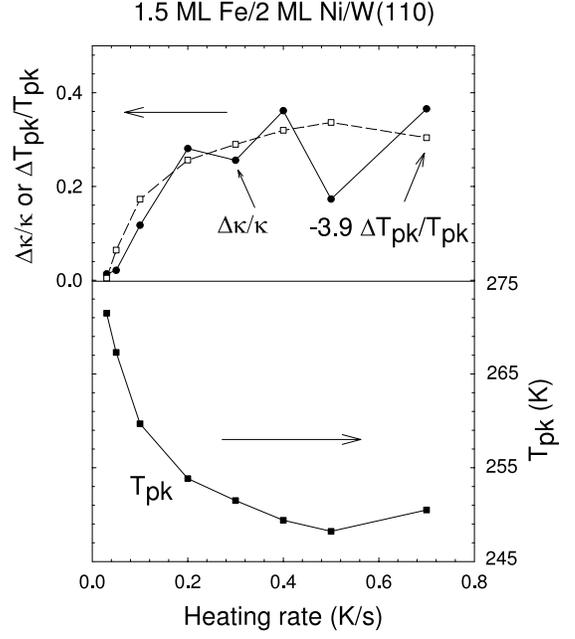}} 
\caption{The bottom portion shows the dependence of the peak temperature of the susceptibility on the heating rate for 1.5ML Fe/2ML Ni/W(110) films, using the right hand scale.  The top part shows both the relative change in the parameter $\kappa$ (solid symbols) and the peak temperature (open symbols), with heating rate, using the left hand scale.}
\end{figure}

The measured shift of the susceptibility peak with heating rate is presented in fig. 2 for the films with 1.5 ML Fe (right hand scale).  According to eq.(\ref{shift}), changes in $T_{pk}$ should be linearly related to the changes in $\kappa$. This is tested, using the left hand scale of fig. 2, by overplotting the relative change in $\kappa$ and in $T_{pk}$. The scaling factor of $\approx -4$ between the two curves is smaller than predicted by eq.(\ref{shift}), which gives a factor of $\approx -32$ for the experimentally determined parameters.  However, the data in fig. 1 and 2 are in qualitative agreement with changes in the susceptibility predicted by the model of a relaxing density of topological defects. 
\begin{figure} 
\scalebox{.4}{\includegraphics{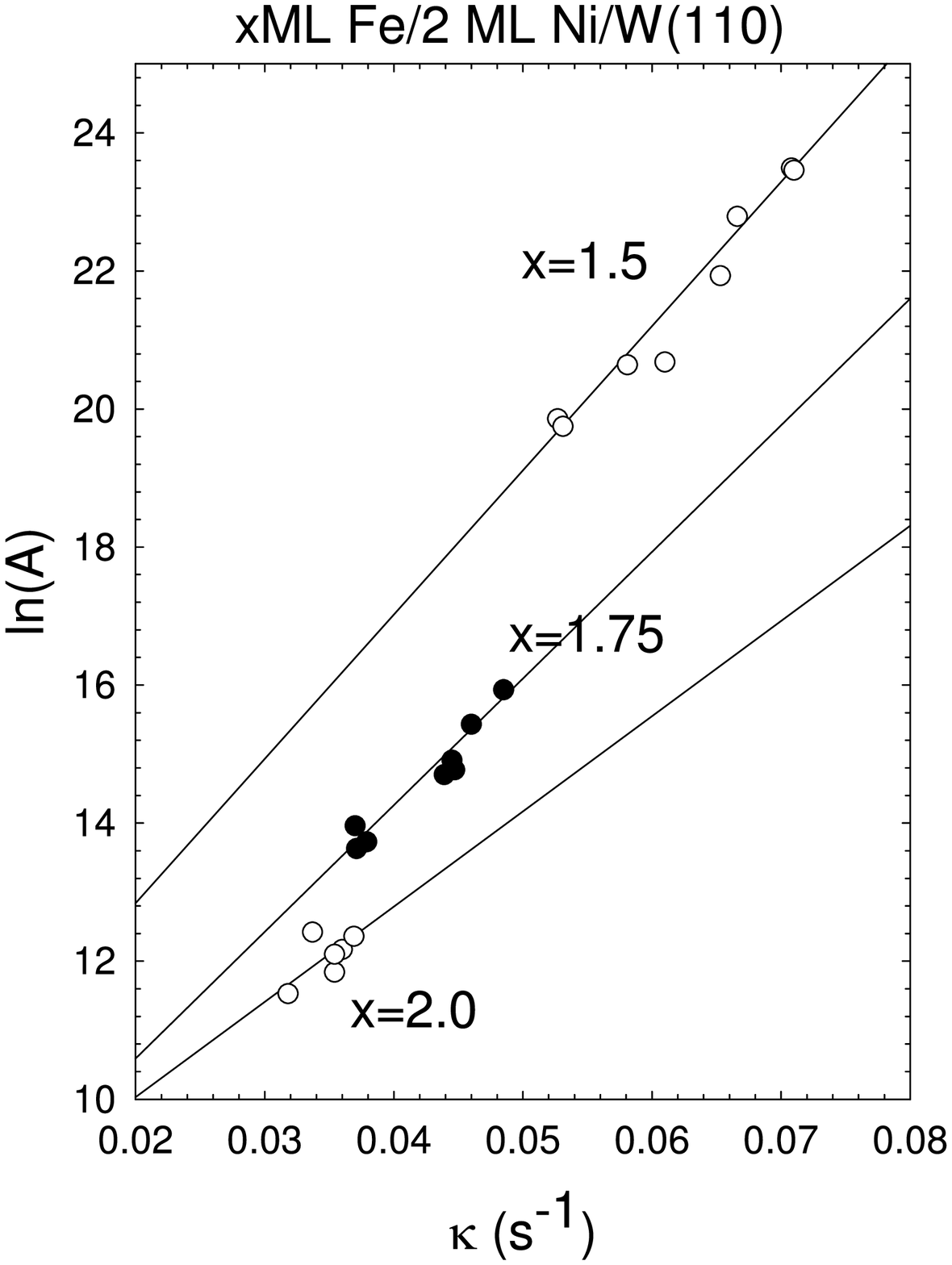}} 
\caption{The fitted parameters $\kappa$ and $\ln A$ from fig. 1 are plotted against each other.  The lines are from a single fit of all the data to eq.(\ref{slopes}).}
\end{figure}

The data permit two quantitative tests of this conclusion.  First, the relations in eq.(\ref{dk}) and (\ref{dA}) are used to test whether the domain energy density changes in a way consistent with the presence of topological defects.  The fitted values of $\ln A$ and $\kappa$ from fig. 1 are plotted in fig. 3.   As predicted, these parameters are linearly correlated.  The lines through the data are fitted according to the prediction in eq.(\ref{slope}). The temperature dependence of the surface anisotropy in eq.(\ref{Keff}) is assumed to be linear.  This has proven to be a good approximation in previous microscopy studies of the stripe width.\cite{won}  Then
\begin{equation}
K_{eff}=\frac{K_{s,0}-\lambda T}{N}-\frac{\Omega}{b},
\end{equation}
and the slopes for the films with different thickness, $N$, in fig. 3 are given by
\begin{equation}
\label{slopes}
m(N)=2[(\frac{K_{s,0}}{\lambda})-T_0-N(\frac{\Omega}{b\lambda})].
\end{equation}
The lines in fig. 3 are a simultaneous least squares fit of all the data that yield the parameters $\frac{K_{s,0}}{\lambda}=900\pm 150$ K and $\frac{\Omega}{b\lambda}=150\pm 40$ K. $T_0$ is taken as $T_{pk}$ for the slowest heating rate for each film thickness.  $N$ is the total film thickness, including the 2 ML Ni. The three intercepts are fit independently of each other, as the absolute magnitude of the susceptibility is expected to vary somewhat from film to film.  The ratio of anisotropy to dipole energy, $\frac{K_{s,0}}{\Omega/b}\approx 6$, is reasonable.  The excellent description of the data confirms that there is an additional contribution to the domain pattern energy that scales with the domain wall energy and $L$, and which changes $\kappa$.  This is consistent with topological defects.  As is shown in the appendix, contributions with different scaling properties give very different results.

The second quantitative test compares the experimental change in $\kappa$ with the heating rate, to that expected for the population dynamics of topological defects. $\Delta\kappa(T)/\kappa_0$ was integrated numerically using eq.(\ref{diffeqk}).  Because fig. 2 comfirms a linear relation between the fractional change in $T_{pk}$ and $\kappa$, the relaxation time $\tau_{an}$ for annihilation of topological defects is equal to the experimentally determined relaxation time for the shift of the susceptibility peak; that is, the fundamental time scale $\tau_{0,an}=0.7$s and the activation energy $E_{an}=1560$K are taken directly from the experiment.  Figure 4 shows calculated curves for $\kappa(T,R)$ for a selection of heating rates.  The input parameters $\kappa_0=0.048$ and $\gamma \epsilon \approx \frac{\kappa_{max}}{\kappa_0}-1$ are taken from fig. 1d.  Figure 4a shows curves generated for two very different sets of initial conditions.  $q_0=0$ corresponds to heating from a state where there are no topological defects, whereas $q_0=2\epsilon$ corresponds to an initial density of defects that is twice the steady state value in the absence of annihiliation.  As can be seen, the initial concentration of topological defects is rapidly diluted by the density of new defects generated by the exponential growth of domain density.
\begin{figure} 
\scalebox{.4}{\includegraphics{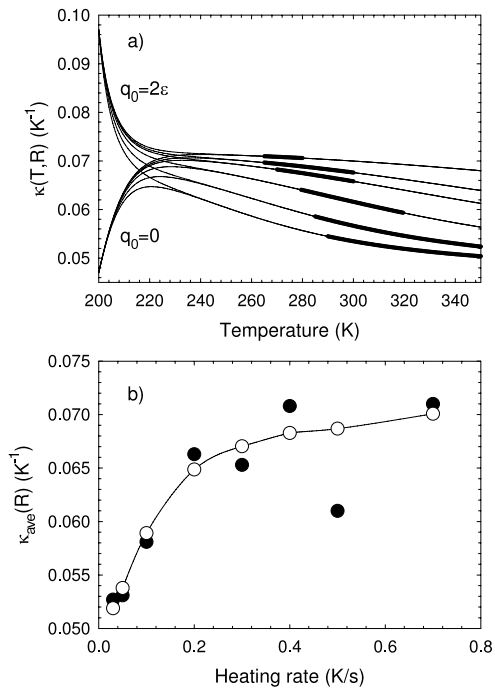}} 
\caption{a) $\kappa (T,R)$ is integrated from eq.(\ref{diffeqk}) using two different initial boundary conditions.  The heating rates, in K/s, are, from bottom to top: 0.03, 0.05, 0.10, 0.20, 0.30, 0.40, 0.70. The thick portion of each curve indicates the temperature range used to fit the value of $\kappa$ in each experimental susceptibility in fig. 1. b) The average value of $\kappa$ within the thick region of the curves in part a), are plotted as open symbols.  The experimentally fitted values from fig. 1 are plotted as closed symbols.}
\end{figure}

The range of temperature over which experimental susceptibility data can be fit to determine the value of $\kappa$ is bounded at low temperature by $T_{pk}$, where pinning becomes important, and at high temperature by the vanishing signal (see fig. 1a).  As $R$ and $\kappa$ increase, this range gets smaller.  The bold regions of the curves in fig. 4a indicate the experimental temperature range used in fitting for the values of $\kappa$ in fig 1d.  The average calculated value of $\kappa (T)$ within this range in fig. 4a, for each value of $R$, is plotted in fig. 4b as the open symbols connected by a smooth curve.  The standard deviation of the calculated value within each range is smaller than the symbol size.  The solid symbols are the experimental points from fig. 1d.  The calculated dynamics of the density of topological defects represents the experimental data for the change in $\kappa$ very well, with no adjustable parameters.

The systematic variation of the experimental susceptibility with heating rate is entirely consistent with the presence and dynamics of a density of metastable topological defects whose energy scales with the domain wall energy, whose size scales quasi-statically with the domain size, and that are produced on a statistical basis by the growth of new domains.  Since $\kappa$ depends only upon $\gamma\epsilon$, this could be due to a low density of strongly perturbing defects, or a high density of weakly perturbing defects.  For example, the defects could be individual dislocations, or bound dislocation pairs.

The experiments in ref.(\onlinecite{libdeh1}) provide further evidence to help resolve this ambiguity.  The magnetic susceptibility was also measured while cooling, although the range of accessible cooling rates was not large enough to conduct a systematic study.  However, it was clear that the relaxation times for the shift in the susceptibility curve when cooling are at least an order of magnitude longer than while heating.  This suggests that the annihilation of a different type of topological defect predominates during the long term relaxation following domain growth (heating) and domain removal (cooling).  

It has already been argued that, during domain growth, dislocations at the end of existing semi-infinite stripes advance toward each other, driven by an existing domain density that is less than the equilibrium value.  The dislocations are effectively attractive and form bound dislocation pairs with a probability $\epsilon$.  During cooling, the domain density must be exponentially reduced, and domain loss is driven by a domain density that is greater than the equilibrium value.  In this case, dislocations retreat from one another, such that their interaction is effectively repulsive.  This acts to unbind dislocation pairs leaving isolated dislocations.  The time scale for the annihilation of unbound dislocations should be much greater than for bound pairs, since the isolated dislocations must first ``find one another''.  This leads to extremely long times for relaxation to the ground state.  The  experimentally observed asymmetry in the relaxation after heating and cooling therefore suggests that it is the population dynamics of bound dislocation pairs that drives the relaxation of the susceptibility curve while heating. 

\section{Conclusions}

Topological defects in the magnetic domain pattern of perpendicularly magnetized films represent the elementary pattern excitations that permit the system to evolve towards equilibrium.  While these defects have often been observed in microscopy studies, their crucial role in the dynamics of pattern formation has been largely unexplored because of the temporal restrictions on imaging techniques, and because it is difficult to follow a large population of the defects using a local probe.  However, the presence of topological pattern defects has a pronounced effect on the magnetic susceptibility arising from domain wall motion. The leading (high temperature) edge of the susceptibility becomes exponentially steeper, and the curve as a whole is shifted to lower temperature.  The shift can be parameterized as correlated changes in the parameters $\kappa$ and $\ln A$, which are related by the temperature dependence of the surface anisotropy.  As the population of defects annihilates on a time scale of order minutes, the whole susceptibility curve relaxes to higher temperature.  This provides a sensitive method of detecting these defects and studying their population dynamics.

These predictions are confirmed quantitatively in recently published experimental results.  The experiments give quantitative support to the following aspects of the model.  During heating, the domain density increases exponentially, and bound dislocation pairs are produced by the stripe growth process as individual dislocations grow towards each other.  The rate of production is proportional to the density of magnetic stripe domains.  The bound pairs are a metastable configuration that annihilates with a long time constant. The experimental activated relaxation time observed for the shift in the peak temperature of the susceptibility is identified as an experimental measurement of the activated relaxation of the bound dislocations.  Upon cooling, the domain density decreases exponentially, and the dislocation pairs unbind and retreat one from another to remove domain stripes.

These findings illustrate the potential for the experimental investigation of the dynamics of a two dimensional pattern-forming system.  In particular, the stripe domain system is expected to melt at a Kosterlitz-Thouless transition by the proliferation of unbound dislocations.  This study suggests that measurements of the magnetic susceptibility will be a powerful tool for the experimental detection and study of this transition.  It offers the potential to clarify the paths this two dimensional magnetic system follows to paramagnetism in both the presence and absence of a reorientation transition to the in-plane ferromagnetic state.

\section{Acknowledgements}
We are thankful for the continuing technical assistance of M. Kiela.  This work was supported by the Natural Sciences and Engineering Research Council of Canada.

\appendix
\section{}
This appendix treats how defect geometries that do not scale with the stripe width $L$ affect the magnetic susceptibility.  If the fixed size of the defect is $L_0$, then the integrals defined in eq.(\ref{comp}) and (\ref{curve}) become $\alpha^0(L_0)\rightarrow L^2_0\alpha^0$ and $\beta^0(L_0)\rightarrow \beta^0$.  The area of the defect is $L^2_0$, so that $f=(QL^2_0)/(L_xL_y)$ is the fraction of the area occupied by this type of defect.  Substituting these into eq.(\ref{Qdyn}), the areal domain energy density becomes
\begin{equation}
\label{newE}
E= [(1+f\gamma_{\alpha})n +f\gamma_{\beta}\frac{1}{L^2_0 n}]E_WN-4\Omega N^2 n \ln[\frac{2}{\pi \ell n}],
\end{equation}
where $\gamma_{\alpha}$ and $\gamma_{\beta}$ are the dimensionless proportionality constants between the compression and curvature energies, respectively, and $E_W$.  The metastable domain density is then
\begin{equation}
n_{ms}(T)= \frac{2}{\pi \ell}\exp[-(1+f\gamma_{\alpha}-f\gamma_{\beta}\frac{1}{L^2_0 n^2_{ms}})\frac{E_W(T)}{4\Omega N}-1].
\end{equation}
\begin{figure} 
\scalebox{.4}{\includegraphics{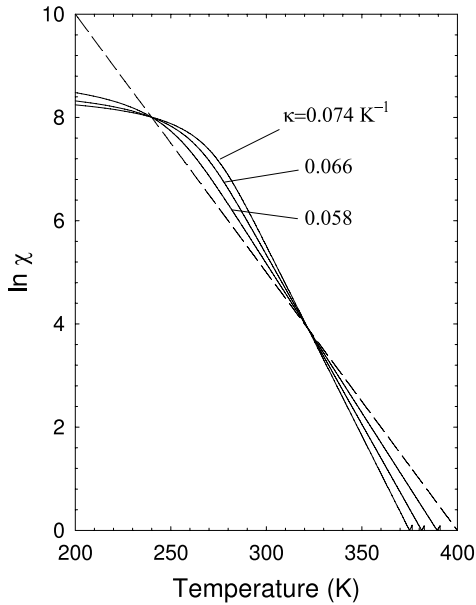}} 
\caption{The susceptibilities in the presence of topological defects of fixed size $L_0$ derived from eq.(\ref{chieq2})-(\ref{T2}) are plotted against temperature.  The dashed line is the case where the defects are absent.  The three solid lines for increasing values of $\kappa$ at high temperature, are for an increasing density of these defects.}
\end{figure}
Comparing to eq.(\ref{nmeta}), this can be written as
\begin{equation}
\ln(n_{ms})=(1+f\gamma_{\alpha}-f\gamma_{\beta}\frac{1}{L^2_0 n^2_{ms}})\ln(n_{eq}).
\end{equation}
By identifying $y=\ln(\pi\ell n_{eq}/2)$ and $x=\ln(\pi\ell n_{ms}/2)$, this is equivalent to
\begin{equation}
y(x)=\frac{x}{1+f(\gamma_{\alpha}-\gamma_{\beta}\frac{\pi^2 \ell^2}{4L^2_0}e^{-2x})}.
\end{equation}
In fig. 5, the solutions
\begin{equation}
\label{chieq2}
\ln[\chi(x)]=\ln(\frac{\ell}{\pi d}) - x
\end{equation}
and
\begin{equation}
\ln[\chi_{eq}(y(x))]=\ln(\frac{\ell}{\pi d}) - y(x)\equiv \ln A_0 -\kappa_0 T
\end{equation}
are plotted against the temperature
\begin{equation}
\label{T2}
T(y(x))=\frac{\ln A_0 -\ln(\frac{\ell}{\pi d})+y(x)}{\kappa_0}.
\end{equation}
For the calculation in the figure, $\ln A_0$ and $\kappa_0$ are taken from fig. 1d and 1e for 1.5 ML Fe;  $\ell/(\pi d)=60$ and $(4L^2_0)/(\pi^2 \ell^2)=3600$.  This gives the defect a dimension $L_0$ that is ten stripes wide, with each stripe having a width of about ten domain walls. For simplicity, $\gamma_{\alpha}=\gamma_{\beta}=\gamma$.  The dashed line gives $\ln(\chi_{eq})$ when there are no defects and $f\gamma =0$.  The three solid lines are for increasing values of $f\gamma$ that produce logarithmic slopes at high temperature of $\kappa=$ 0.058, 0.066 and 0.074, covering the range observed in fig. 1c and fig. 4.

An important difference between fig. 5 and the analysis in eqs.(\ref{neq}-\ref{dA}) is that the susceptibility saturates quickly when $L\sim L_0$ as a result of the fixed scale of the defect.  This is because the curvature energy in eq.(\ref{newE}), that is proportional to $1/n$, grows very quickly when the radius of curvature approaches the stripe width.  This in turn limits the reduction of domain density as the temperature is reduced.

The traces in fig. 5 show that, as the density $f$ of defects is reduced, $\kappa$ decreases and the onset of domain saturation moves to lower temperature. While the former effect is seen in the experiment, the latter would move the peak temperature to lower temperature when the heating rate decreased (i.e. the integrated elapsed time increased).  This is opposite to what is observed in the experiments (see fig. 2).  Furthermore, if the point of saturation changed in this way as the density of defects was reduced, then the fitted values for the pinning parameters in fig. 1b and 1c would change with the heating rate.  These effects would become even more pronouced for smaller defects.  For these reasons, it is unlikely that topological defects, or structural defects, that do not scale quasi-statically with the domain width are responsible for the slow relaxation of the susceptibility.


\begin{thebibliography}{39}
\bibitem{chaikin}P.M. Chaikin and T.C. Lubensky, \emph{Principles of condensed matter physics}, (Cambridge University Press, Cambridge, 2000).
\bibitem{bander}M. Bander and D.L. Mills, Phys. Rev. B \textbf{38}, 12015 (1988).
\bibitem{sagui}C. Sagui and R.C. Desai, Phys. Rev. Lett. 
\textbf{71}, 3995 (1993); Phys. Rev. E \textbf{52}, 2807 (1995). 
\bibitem{sikes}H.D. Sikes and D.K. Schwartz, Science, 1604 (1997).
\bibitem{vanderbilt}D. Vanderbilt, Surface Science Lett. \textbf{268}, L300 (1992).
\bibitem{kivelson}S.A. Kivelson, I.P. Bindless, E. Frandkin,
V. Oganesyan, J.M. Tranquada, A. Kapitulnik, and C. Howard,
Rev. Mod. Phys. \textbf{75}, 1201 (2003).
\bibitem{skomski}R. Skomski, H.-P. Oepen and J. Kirschner, Phys. Rev. B \textbf{58}, 3223 (1998); Y. Yafet and E.M. Gyorgy, Phys. Rev. B \textbf{38}, 9145 (1988).
\bibitem{allenspach}R. Allenspach and A. Bischof, Phys. Rev. Lett. \textbf{69}, 3385 (1992).
\bibitem{stoycheva}A.D. Stoycheva and S.J. Singer, Phys. Rev. Lett. \textbf{84}, 4657 (2000);  J.P. Whitehead, A.B. MacIsaac and K. De'Bell, Phys. Rev. B \textbf{77}, 174415 (2008).
\bibitem{vaterlaus}A. Vaterlaus, C. Stamm, U. Maier, M.G. Pini,
P. Politi, and D. Pescia, Phys. Rev. Lett. \textbf{84}, 2247 (2000).
\bibitem{won}C. Won, Y.Z. Wu, J. Choi, W. Kim, A. Scholl, A. Doran, T. Owens, J. Wu, X.F. Jin, H.W. Zhoa, and Z.Q. Qui, Phys. Rev. B \textbf{71}, 224429 (2005).
\bibitem{kosterlitz}J.M. Kosterlitz and D.J. Thouless, J. Phys. C\textbf{6}, 1181, (1973);  D.R. Nelson and B.I. Halperin, Phys. Rev. B \textbf{19}, 2457 (1979).
\bibitem{kashuba}A.B. Kashuba and V.L. Pokrovsky, Phys. Rev. B
\textbf{48}, 10335 (1993).
\bibitem{abanov}A. Abanov, V. Kalatsky, V.L. Pokrovsky, and
W.M. Saslow, Phys. Rev. B \textbf{51}, 1023 (1995).
\bibitem{saratz}N. Saratz, T. Michlmayr, O. Portmann, U. Ramsberger, A. Vaterlaus and D. Pescia, J. Phys. D \textbf{40}, 1268 (2007).
\bibitem{portmann}O. Portmann, A. Vaterlaus, and D. Pescia, Nature (London) \textbf{422}, 701 (2003)
\bibitem{bromley}S.P. Bromley, J.P. Whitehead, K. De'Bell and
A.B. MacIsaac, J. Magn. Magn. Mater. \textbf{264}, 14 (2003).
\bibitem{cannas}S.A. Cannas, M.F. Michelon, D.A. Stariolo and
F.A. Tamarit, Phys. Rev. E \textbf{78}, 051602 (2008).
\bibitem{portmann2}O. Portmann, A. Vaterlaus, and D. Pescia, Phys. Rev. Lett. \textbf{96}, 047212 (2006).
\bibitem{libdeh1}N. Abu-Libdeh and D. Venus, Phys. Rev. B
\textbf{80}, 184412 (2009).
\bibitem{libdeh2}N. Abu-Libdeh and D. Venus, Phys. Rev. B \textbf{81}, 195416 (2010).
\bibitem{chikazumi}S. Chikazumi, \emph{Physics of Ferromagnetism}, (Clarendon, Oxford, 1997).
\bibitem{venus1}D. Venus, C.S. Arnold and M. Dunlavy, Phys. Rev. B\textbf{60}, 9607 (1999).
\bibitem{venus2}D. Venus and M.J. Dunlavy,
J. Magn. Magn. Mater. \textbf{260}, 195 (2003).
\bibitem{darby}M.I. Darby and B.K. Middleton, J. Phys. D: Appl. Phys. \textbf{6}, 116 (1973).
\end{thebibliography}
\end{document}